**Daniel Bernoulli's Research in Basel and the "Physikalisches Kabinett" in the "Stachelschützenhaus"**

On 22 September 2023 the inauguration of the "Stachelschützenhaus" at Basel's Petersplatz (St. Peter's Square) as an EPS Historic Site was hosted by the University of Basel. In the following article we will focus first on Daniel Bernoulli's career path, before discussing his major scientific achievements and finally adding some aspects of the inauguration ceremony (the colloquium talks, the visit of today's laboratories in the Stachelschützenhaus, and the unveiling of the plaque).

Daniel Bernoulli (1700-1782, a member of the world-renowned Bernoulli family of mathematicians and scientists that had been based in Basel since 1623) studied medicine in Basel, Heidelberg and Strasbourg.

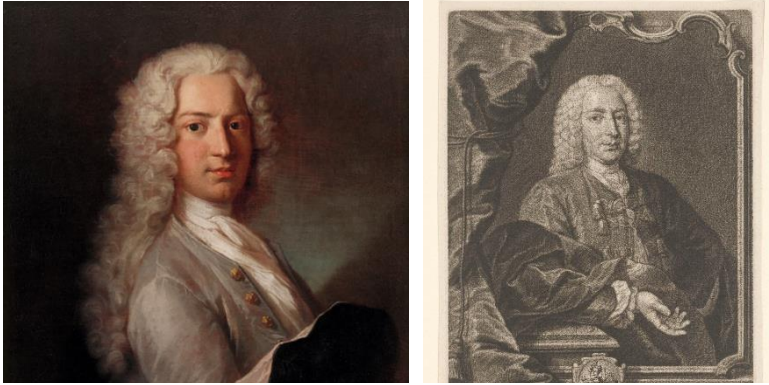

Fig.1: portraits of Daniel Bernoulli (a) Basel, ca. 1720–25 (Historisches Museum Basel), (b) by Johann Rudolf Huber (1744) (wikimedia)

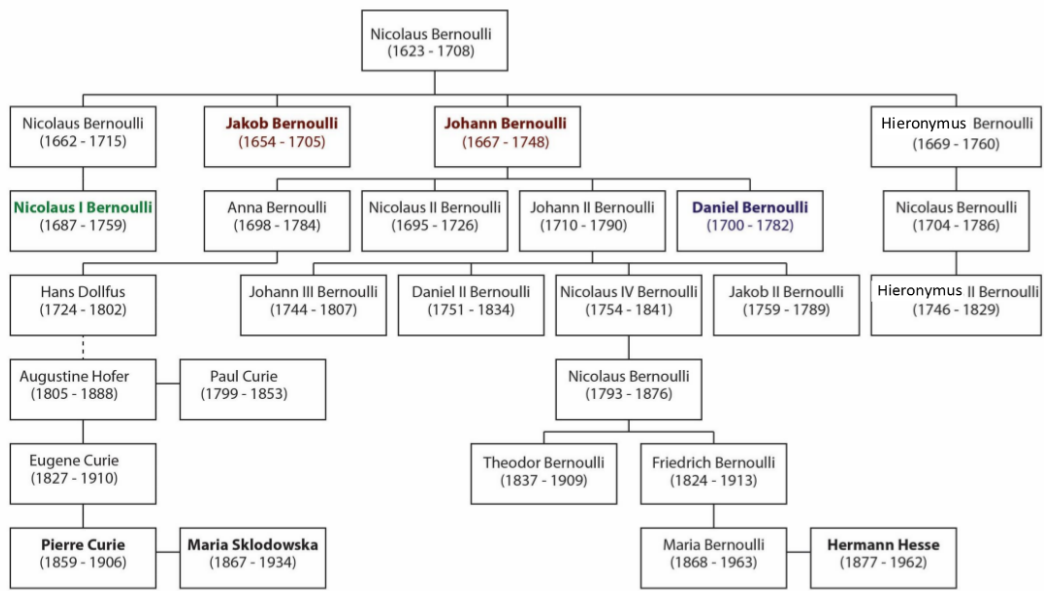

Fig.2: (Part of) The Bernoulli Family Genealogy (according to https://galileo-unbound.blog/2020/10/06/the-bountiful-bernoulli-of-basel/). Daniel Bernoulli was born in Groningen in 1700, where his father Johann (I) Bernoulli then held a chair. After his uncle Jacob (I) had died in 1705, the family returned to Basel.

**Daniel Bernoulli's career:** He obtained his doctoral degree in 1721 with a thesis on the mechanics of respiration. He was the first in history to approach this topic in a mathematical way. Utilizing the methods of exact sciences, also in applications for solving problems in physiology and engineering, was to become the trademark for his entire career in research. In 1724 he published his first little book, "Exercitationes", which treated among other things the Riccati differential equation. This attracted so much attention that in 1725 he was appointed to the newly founded Academy at St. Petersburg, jointly with his brother Nicolaus II, a gifted mathematician.

After his brother's early death in 1726 he began planning to return to Basel, but was only able to do so when a chair – actually in anatomy and botany – became vacant in 1733. Only in 1750 he would become professor of physics at the University of Basel. From then on until 1776, he gave remarkable physics lectures, including experiment demonstrations, and was highly active in his favorite research field.

He published studies related to medical aspects (such as blood circulation, cardiac work, medical statistics, and epidemiology) and applications of mathematical physics to practical problems in mechanical and nautical engineering, but his main dedication was to physics problems with a strong mathematical footing.

His overall scientific oeuvre includes 74 papers and he won no less than ten of the yearly prizes of the Paris Académie des Sciences for topics in astronomy, physics and its applications to nautical engineering: the determination of positions at sea by various timekeeping devices (1725, 1745-47) and compasses (1743), the geometry of the solar system (1734), the optimal shape of anchors (1737), the theory of the tides (1740), magnetism (1744-46), ocean currents (1751), the propulsion of ships (1753), and the reaction of their pitching and rolling motion (1757). Many of these aspects can be characterized by his approach to solve physics problems via being a pioneer in the development of mathematical physics, using the powerful calculus of Leibniz in Newton's theories! A more detailed account of his scientific work in general as well as at the University of Basel is given e.g. in refs. [1-5].

**Main scientific achievements:** His most comprehensive work, the "Hydrodynamica" (authored mainly in St. Petersburg, completed and announced in 1734, but finally printed at Strasbourg only in 1738) achieved a fundamental advance in fluid mechanics and laid the foundations for later progress. The important distinction between hydrostatic and hydrodynamic pressure is due to him. He formulated a special form of the "Bernoulli equation" for incompressible and frictionless fluids, treated elastic fluids for the first time and marked the beginnings of the kinetic theory of gases.

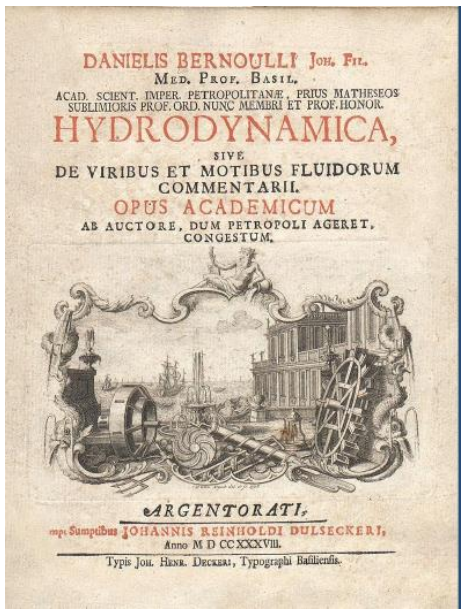

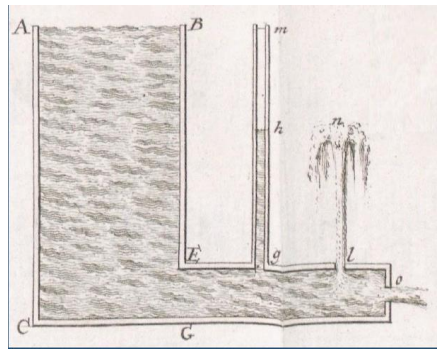

Fig.3: Original copy of Bernouli's "Hydrodynamica" printed by Johann Reinhold Dulsecker (Strasbourg) in 1738 [6]

Fig.4: Pressure reduction due to Bernoulli's equation "an increase in speed of a flowing liquid leads to a decrease in its pressure" (in the right part where the liquid is moving) [6].

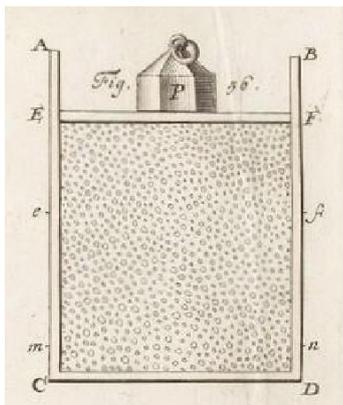

Fig.5 Bernoulli pioneered a molecular model of gases in order to explain the Boyle-Mariotte law. He derived macroscopic properties of a gas from microscopic movements of its particles [6].

In other fields, too, Daniel Bernoulli was always interested in a close connection between mathematical methods and their applications. He investigated special techniques for solving algebraic equations, dealing with continuous fractions, and solving questions of probability theory. He wrote on the composition of forces, their "true measure" and their conservation, analyzed the oscillations of mechanical systems (chains, vibrating strings and blades). In the course of a debate with Euler and d'Alembert on the vibrations of a string, his physical intuition prevailed, leading to an early form of what would become Fourier analysis. He made fundamental contributions to the theory of collisions, the theory of tides, winds and marine currents, the longitude problem (with the aid of an inclination compass for measuring not only the horizontal component of the Earth's magnetic field but also the perpendicular component to the earth's surface at a given location), as well as the geometry and dynamics of the solar system. Bernoulli also undertook a determination of the work of the human heart (arriving at about 0.6W compared to today's values of 1 to 1.5W) and developed a statistical model for epidemics (in the context of smallpox inoculation).

Bernoulli was the first to point out the decomposition of the motion of bodies into translational and rotational motions (a structure that resembles Lagrange's Analytical Mechanics, since all results appear as a consequence of a single principle, in this case the conservation of energy). The above-mentioned Bernoulli's law of flow (the energy theorem for stationary flows) is today the general basis for fluid dynamics (hydrodynamics and gas dynamics) and thus also for the technology of aviation. Bernoulli was able to sketch (although not in full detail) the equation of state analyzed by Van der Waals a century later. Based on experimental evidence, he was also able to conjecture certain laws which were not verified until many years later; among them was Coulomb's law in electrostatics (see again [1-5]).

**The South Wing of the Stachelschützenhaus, where the Physics Cabinet (Physikalisches Kabinett) was located in the 18th century**

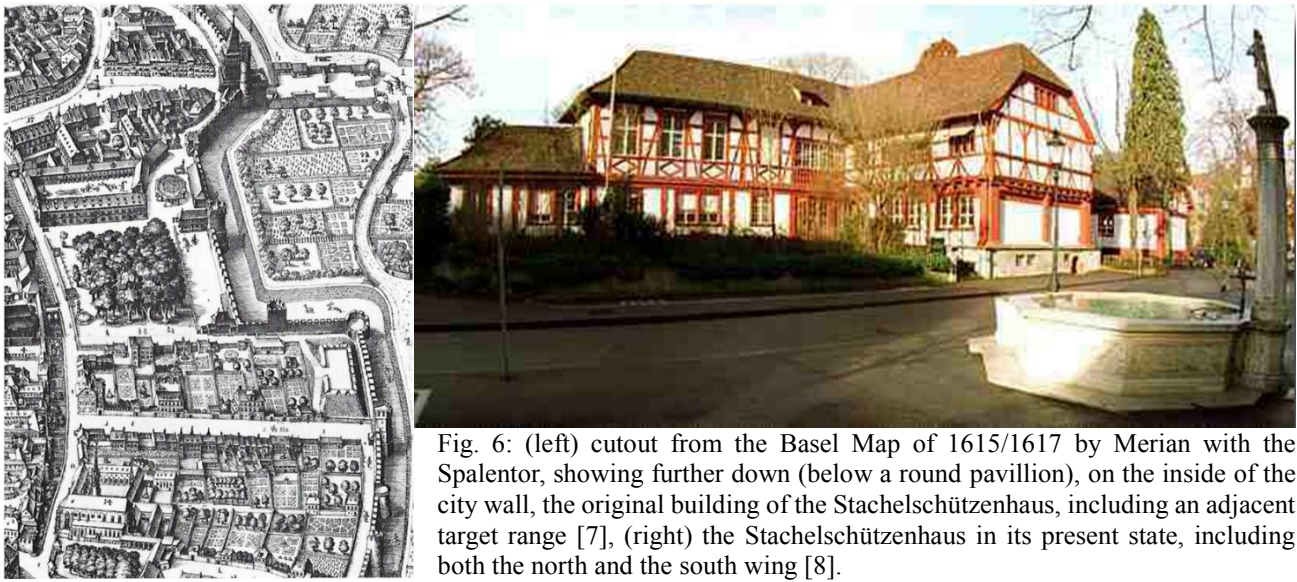

Fig. 6: (left) cutout from the Basel Map of 1615/1617 by Merian with the Spalentor, showing further down (below a round pavillion), on the inside of the city wall, the original building of the Stachelschützenhaus, including an adjacent target range [7], (right) the Stachelschützenhaus in its present state, including both the north and the south wing [8].

The Stachelschützenhaus was initially erected in 1519/20 (other sources give 1546) [8-10] and served for training the municipal Crossbow Guard. In 1709 a first extension was built to the north, in 1729 a second wing to the south, in order to house the University's collection of physical instruments. The professor of physics and botany at that time, Benedict Staehelin (1695–1750), started a collection of physical devices and instruments that he had acquired for demonstration purposes. Optical, pneumatic, and mechanical appliances were ordered from the British instrument maker Francis Hawksbee. In 1747, after some rumors about neglect of the laboratory by Staehelin, due to illness, the mayor requested a report from the medical faculty about the status of the collection. The latter was undertaken by Daniel Bernoulli, leading to the request of more money for the necessary improvements and an assistant position to keep it in optimal shape. After Bernoulli became professor of physics in 1750, he greatly expanded the Physics Cabinet by adding many instruments for his research, a 1752 inventory includes more than 130 items. Additional acquisitions with a value of several hundred pounds extended the inventory to a list of 40 pages in 1757.

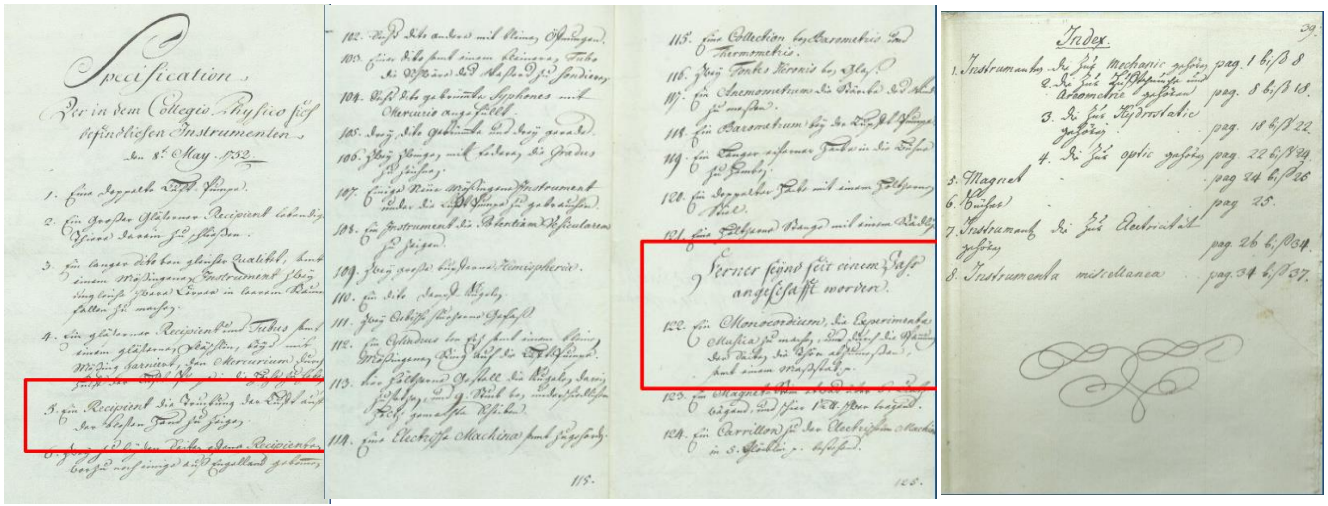

Fig.7: Excerpts of Bernoulli's 1752 inventory of the Physics Cabinet: (left) indicating (via red borderlines) a water barometer with the water level inside recording the air pressure, (middle) a monochord, demonstrating the change of the sound frequency with the changing tension of cords, (right) the overall index of all devices (from the Basel State Archive {11})

As a professor of Physics Daniel Bernoulli performed his experimental activities in the "Physikalisches Kabinett" (Physics Cabinet) in the south wing of the Stachelschützenhaus. Some of the many instruments of his Physics Cabinet are still in existence today. Among them is the experiment for the demonstration of the "hydrostatic paradoxon" which vividly demonstrates that the pressure in a liquid is independent of the shape of the vessel and depends only on the height of the liquid column. Some of these early experimental devices are today in the possession of the Historisches Museum Basel and on display in the Haus zum Kirschgarten, where one finds instruments and experiments that were acquired by Daniel Bernoulli and his predecessor Benedikt Staehelin.

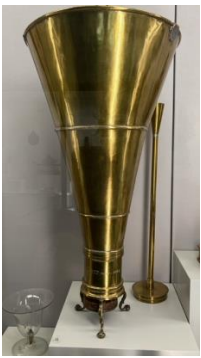

Fig. 8: "Hydrostatic Paradoxon" Device for demonstrating the Hydrostatic Paradox, with the following description: A wide and a narrow vessel are screwed on the same cylinder. If the water column is the same, the bottom pressure in both vessels is the same. This hydrostatic paradox states that the gravitational pressure exerted by a liquid in a vessel on the bottom of the vessel, while dependent on the level of the liquid, is, on the other hand, independent of the shape of the vessel and thus of the quantity of liquid contained - an example of the fundamental insights of Daniel Bernoulli. Manufacturer: Johann Ulrich Goetz (1694-1758), Basel, 1728, since 1757 in the collection of the Institute of Physics of the Univ. of Basel (photo Salome Noah, Basel).

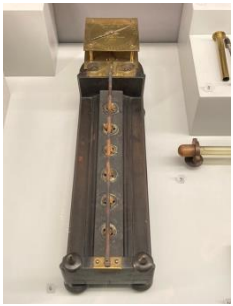

Fig. 9: Musschenbroek's pyrometer (thermal expansion meter), as an example for Bernoulli's lectures with experiment demonstrations: The device was used to demonstrate the expansion of metal under heat. In this process, an iron rod is heated by six spiritus flames and its expansion or coefficient of expansion is indicated by a special transmission mechanism with a pointer on a dial. The pyrometer of this type was invented in 1731 by the Dutchman Pieter van Musschenbroek (1692-1761). Made by Jean Pierre Charme, Paris, c. 1752, iron, brass, wood; purchased by Prof. Daniel Bernoulli (photo Salome Noah, Basel).

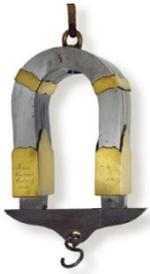

Fig.10: One of the first horseshoe magnets ever made, carrying more than 30 kg, made by the goldsmith Johann Dietrich in 1755, who built several instruments for the Cabinet (photo Daniel Suter, Basel).

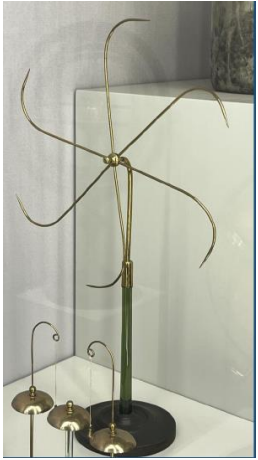

Fig.11: A "sparkling wheel" which was set in rotation by emitting sparks from the endpoint of the curved needles and a carillon with tinkling little bells rung by charged pendulums (photo Daniel Suter, Basel).

Figs. 8-11 show objects at display in the Haus zum Kirschgarten of the Historisches Museum Basel [12]. In the lecture hall of the Physics Cabinet, Bernoulli held lectures with experimental demonstrations over a wide variety in the subfields of physics for students as well as the general public, which enjoyed great popularity for a quarter of a century. This included also aspects of "entertainment physics" (see Fig. 11). After Daniel Bernoulli's death in 1782 the chair of physics was given to a physician, Johann Jakob Thurneisen the Younger (1756-1804), who showed little inclination to use the cabinet, so that the premises fell into disrepair. Thereafter the Stachelschützenhaus has been utilized for quite a number of purposes; presently it is the domicile of the Institute for Medical Microbiology at the University of Basel.

**The inauguration of the Stachelschützenhaus as an EPS Historics Site**
The importance of the laboratory of Daniel Bernoulli, where he performed his ground-breaking research as a true pioneer in the field of physics in Switzerland, Europe, and world-wide has been the main aim to propose the "Stachelschützenhaus" as a Historic Site of the European Physical Society (EPS) in order to serve as a truly important place for the historical development of physics [13]. The proposal was accepted and on 22 September 2023 the inauguration took place in form of a twofold event: As part of the regular Physics Colloquium of the University of Basel the symposium included an introduction to EPS activities and the EPS Historic Sites project by Anne Pawsey (EPS Secretary General), an extended scientific and historical overview by Martin Mattmüller (Bernoulli-Euler Zentrum) about Daniel Bernoulli and his Physics Cabinet (from which much of the text of the above sections is drawn), and finally a review of today's understanding of Astrophysical Gas Dynamics and its relation to the framework of Bernoulli's Theorem in Fluid Dynamics by Stephan Rosswog (University of Hamburg and Stockholm University).

Stephan Rosswog gave a general introduction to fluid dynamics and under which conditions Bernoulli's law actually applies, i.e. flows are stationary, constant entropy of the system, and the acting forces can be derived from a potential ψ. This leads to the fact (with **v·grad** B being a directional derivative and the specific enthalpy h being expressed via the specific energy u and the pressure divided by the density h=u+P/ρ) that the "Bernoulli constant" B is conserved on stream lines. For a vanishing change in the specific energy u and the density ρ this leads - when multiplying by ρ – to the well-known equation ½ ρv² + P +ρgz= const, when utilizing the gravitational potential with height z).

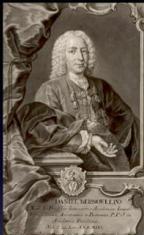

Fig. 12: Deriving the Bernoulli law for specific conditions (from Rosswog's talk, see text).

In an eloquent talk he further discussed details of computational methods in general relativistic hydro-dynamics with the code SPHINCS_BSSN [14], applied to one of the most exciting discoveries in astrophysics: neutron star mergers (first observed in 2017), responsible for the emission of gravitational waves, the ejection of heavy elements up to U and Th, so-called kilonova observations in visible and infrared light, plus a short gamma-ray burst (see Fig.13). This topic has followed him since his PhD in Basel in 1999 [15]. More generally he concluded how fluid dynamics is today a highly active and important research field with a huge range of applications in astrophysics, industry, weather predictions, aviation, special effects in movies etc., and that Bernoulli's law has many applications in every-day life and state-of-the-art astrophysics.

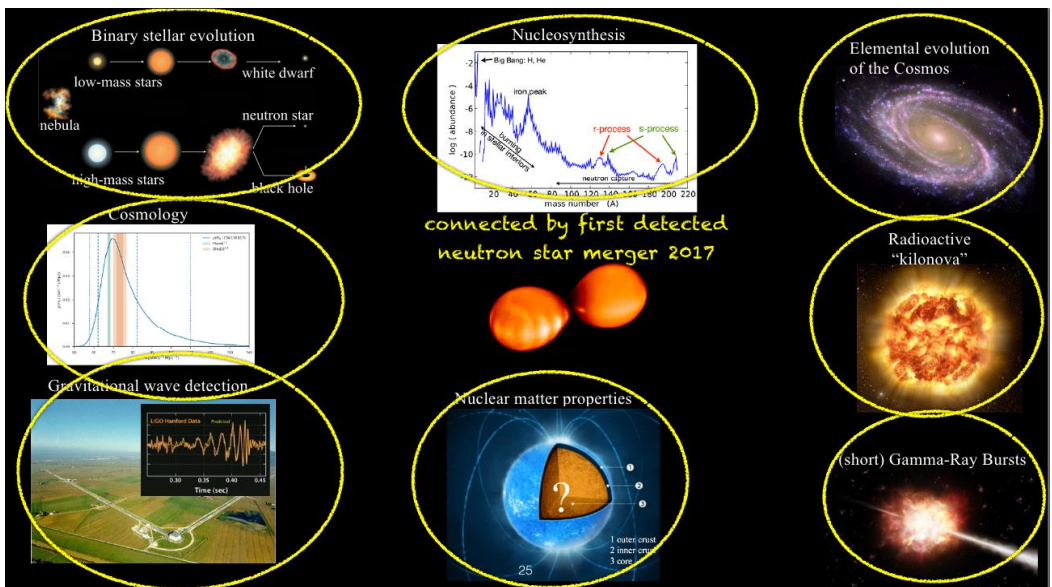

Fig.13: A binary stellar system leads after two supernova explosions to a neutron star binary which, after merging due to gravitational wave emission, causes the ejection of heavy elements and the observation of a kilonova and a short gamma-ray burst. This has a major impact on the evolution of the abundance pattern of heavy elements in our galaxy (from Rosswog's talk).

After Rosswog's talk followed a walk to the Stachelschützenhaus and a short introduction by Rainer Gosert and Klaudia Nägele (Medical Microbiology, Univ. of Basel) about the present activities in clinical virology (the study of viruses and virus-like agents, including their classification, disease-producing properties and genetics) in the Stachelschützenhaus, accompanied by a visit to the present laboratories. Afterwards the EPS Historic Site plaque was unveiled by the EPS General Secretary and the local organizers (Philipp Treutlein, physics department chair, Ernst Meyer, president of the platform MAP within the SCNAT, and Friedrich Thielemann), followed by an aperitif and dinner with enjoyable and enlightening discussions.

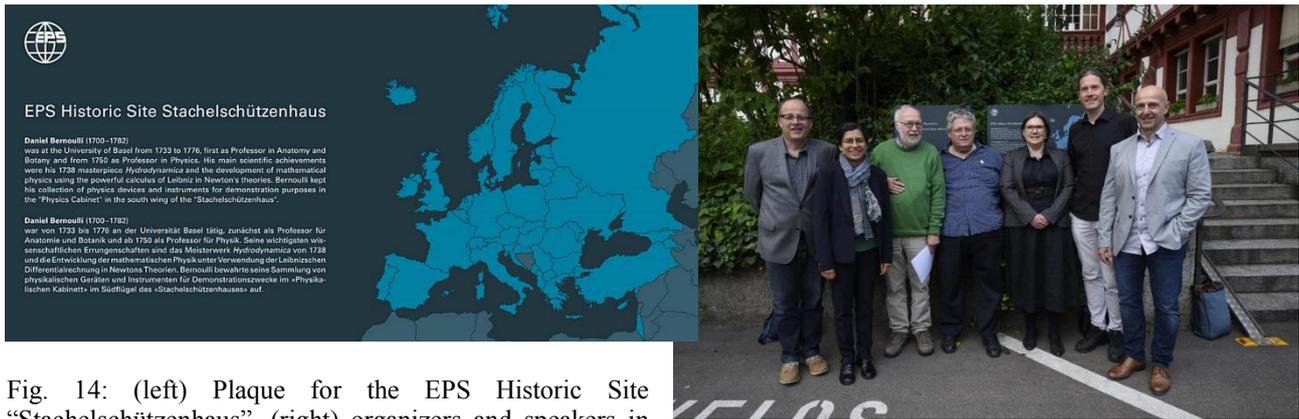

Fig. 14: (left) Plaque for the EPS Historic Site "Stachelschützenhaus", (right) organizers and speakers in front of the plaque and the south wing of the Stachelschützenhaus: Ernst Meyer (president platform MAP/SCNAT and Physics Dept. Univ. of Basel), Gina Gunaratnam (communication coordinator EPS), Friedrich-Karl Thielemann (Physics Department, Univ. Basel), Martin Mattmüller (BEZ, Univ. of Basel), Anne Pawsey (EPS General Secretary), Philipp Treutlein (department chair, Physics Dept. Univ. of Basel), Stephan Rosswog (Univ. of Hamburg and Stockholm Univ.); photos Dominik Plüss, Universität Basel.

We thank Jasmine Brüderlin (Basler Staatsarchiv) and Gudrun Piller (curator of the Haus am Kirschgarten) for the permisions to use Figs.7 and 8-11, respectively, in this article.

The authors: Martin C. E. Huber (Physics Department, ETHZ), Martin Mattmüller (BEZ, Univ. of Basel), Ernst Meyer (president platform MAP/SCNAT and Physics Department, Univ. of Basel) and Friedrich-Karl Thielemann (Physics Department, Univ. of Basel and GSI Helmholtz Center for Heavy Ion Research, Darmstadt)

**References**
[1] Speiser, David, 1982, Daniel Bernoulli (1700-1782), Helv. Phys. Acta 55, 504-523
[2] Die Gesammelten Werke der Mathematiker und Physiker der Familie Bernoulli, eds. Naturforschende Gesellschaft in Basel, Birkhäuser Verlag, contributions about Daniel Bernoulli in volumes 1-8
[3] Dörflinger, Gabriele, 2015, Materialsammlung Daniel Bernoulli, Heidelberger Texte zur Mathematikgeschichte, Universitätsbibliothek Heidelberg
[4] Unigeschichte seit 1460, Departement Physik, https://unigeschichte.unibas.ch/fakultaeten-und-faecher/philnat-fakultaet/zur-geschichte-der-philnat-fakultaet/physik/das-19-jahrhundert-etablierung-der-physik-als-eigenes-fach
[5] Mohr, Anna 2010, Geschichte des Departements Physik, https://unigeschichte.unibas.ch/fileadmin/user_upload/pdf/Mohr_GeschichtePhysik.pdf
[6] e-rara, Plattform für digitalisierte Drucke aus Schweizer Institutionen, https://www.e-rara.ch/
[7] Figure from Matthäus Merian the Elder, Bird's-eye view of Basel from the north-east (copperplate engraving, from Staatsarchiv Basel-Stadt, BILD_1_291).
[8] Unigeschichte seit 1460, Das Stachelschützenhaus
https://unigeschichte.unibas.ch/behausungen-und-orte/universitaetsgebaeude-der-moderne/weitere-gebaeude/das-stachelschuetzenhaus
[9] Bernoulli, Lion, 1980, Geschichte des Stachelschützenhauses in Basel, seperate print from volume 80 of the "Basler Zeitschrift für Geschichte und Altertumskunde", Historische und Antiquarische Gesellschaft zu Basel
[10] Rebmann, Roger Jean, Das Stachelschützenhaus, https://altbasel.ch/haushof/stachelschuetzenhaus.html
[11] Copies of texts from Staatsarchiv Basel-Stadt, Erziehung DD 18.
[12] Haus zum Kirschgarten https://www.hmb.ch/museen/haus-zum-kirschgarten/
[13] EPS Historic Sites, https://www.eps.org/page/distinction_sites
[14] Rosswog, Stephan & Diener, Peter (2021) SPHINCS_BSSN: a general relativistic smooth particle hydrodynamics code for dynamical spacetimes, Classical and Quantum Gravity 38, id.115002, 39
[15] Rosswog, Stephan et al. 1999, Mass Ejection in Neutron Star Mergers, Astron. Astrophys. 341, 499-526; Freiburghaus, Christian et al. 1999, r-Process in Neutron Star Mergers, Astrophys. J. Lett. 525, L121-L124